\documentclass[twocolumn, showpacs,preprintnumbers,amsmath,amssymb, aps]{revtex4}

\usepackage{graphicx}

\begin{document}

\addtolength{\textheight}{1.2cm}
\addtolength{\topmargin}{-0.5cm}

\newcommand{\etal} {{\it et al.}}

\title{Coherently coupling distinct spin ensembles through a high-$T_c$ superconducting resonator}

\author{A. Ghirri,$^{1}\footnote{alberto.ghirri@nano.cnr.it} $ C. Bonizzoni,$^{1,2}$ F. Troiani,$^{1}\footnote{Corresponding author. filippo.troiani@unimore.it} $ N. Buccheri,$^3$ L. Beverina,$^3$ A. Cassinese,$^4$ M. Affronte$^{1,2}$}

\affiliation{$^1$Istituto Nanoscienze - CNR, via Campi 213/a, 411125 Modena, Italy;\\
$^2$Dipartimento FIM, Universit\`a di Modena e Reggio Emilia, via Campi 213/a, 411125 Modena, Italy;\\
$^3$ Universit\`{a} di Milano-Bicocca, Dipartimento di Scienza dei Materiali, Via R. Cozzi 53, 20125 Milano, Italy.\\
$^4$ CNR-SPIN and Dipartimento di Fisica, Universit\`{a} di Napoli Federico II, 80138 Napoli, Italy;\\\\
}

\date{\today}

\begin{abstract}

The problem of coupling multiple spin ensembles through cavity photons is revisited by using PyBTM organic radicals and a high-$T_c$ superconducting coplanar resonator. An exceptionally strong coupling is obtained and up to three spin ensembles are simultaneously coupled. The ensembles are made physically distinguishable by chemically varying the $g$ factor and by exploiting the inhomogeneities of the applied magnetic field. The coherent mixing of the spin and field modes is demonstrated by the observed multiple anticrossing, along with the simulations performed within the input-output formalism, and quantified by suitable entropic measures.

%We demonstrate the coherent coupling between remote ensembles of spin radicals, mediated by confined microwave photons in a high $T_c$ superconducting coplanar resonator. The strong coupling regime between one ensemble of organic magnetic molecules (each carrying an $ s = 1/2$ spin) and the resonator persists up to $30\,$K, and a cooperativity $ C = g^2 / \gamma \kappa $ as high as 4300 is achieved at 2 K, being $g$ the collective coupling strength, $\kappa$ and $\gamma$ the relaxation rates of the cavity photon and of the spin ensemble, respectively. Up to three spin ensembles are simultaneously coupled to the resonator, and are made physically distinguishable from each other by chemically varying the $g$ factors and by exploiting the inhomogeneities of the applied magnetic field. The observed multiple anticrossing, along with the simulations performed within the input-output formalism, demonstrate the hybridization between spin and photon excitations, and thus the coherent coupling between remote ensembles of molecular spins.

\end{abstract}

%03.67.Lx 	Quantum computation architectures and implementations
%33.90.+h 	Other topics in molecular properties and interactions with photons (restricted to new topics in section 33)
%42.50.Pq 	Cavity quantum electrodynamics; micromasers
%76.30.Rn 		Electron Paramagnetic resonance with Free radicals
%75.50.Xx 		Molecular magnets

\maketitle

%%%%%%%%%%%%%%%%%%%%%%
\begin{figure}[ptb]
\begin{center}
\includegraphics[width=8.5cm]{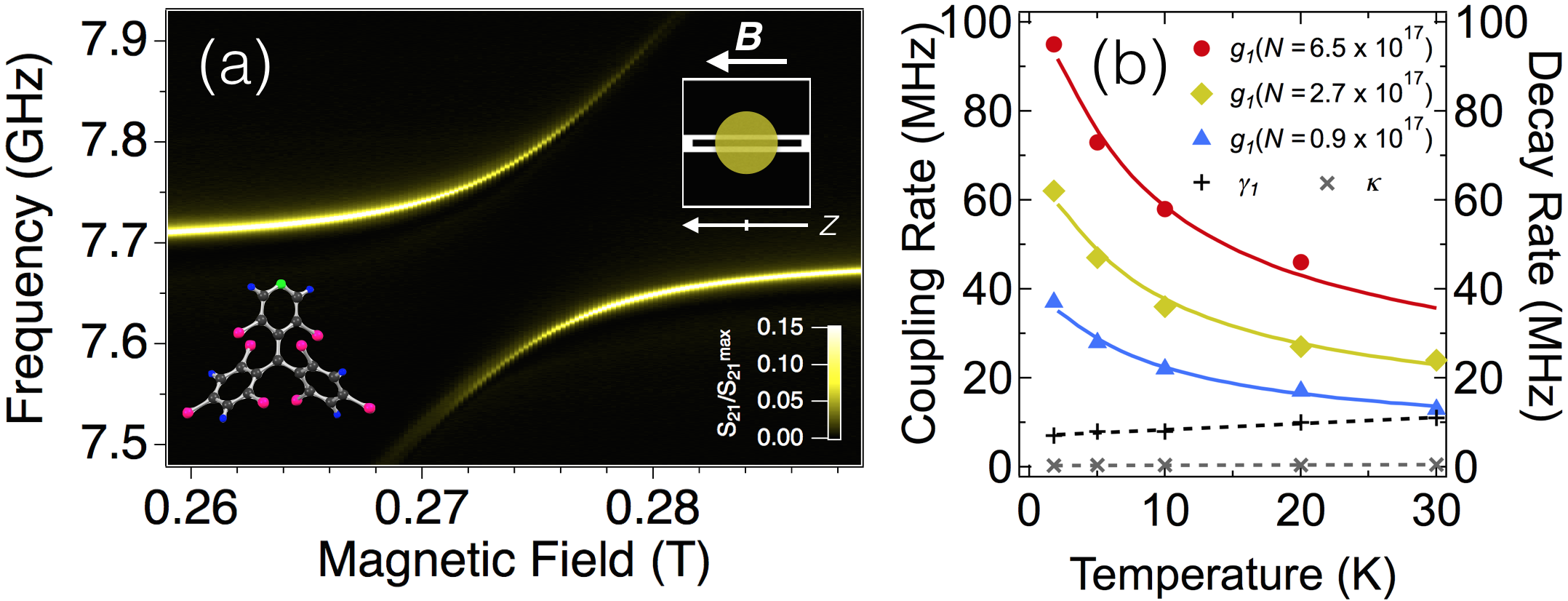}
\end{center}
\caption{(a) Transmission spectra ($T=2\,$K) with a PyBTM ensemble positioned at the center of the YBCO coplanar resonator (right inset). Left inset: molecular structure of the PyBTM molecule: C (black), Cl (magenta), N (green), H (blue). 
(b) Temperature dependence of the decay rates, of the spin-photon coupling (symbols) and of the polarization factor $p(T)$ (solid lines) for different PyBTM ensembles.}
\label{1ens}
\end{figure}
%%%%%%%%%%%%%%%%%%%%%%%

Controlling light-matter interaction at the quantum level is a central problem in modern physics and technology. The paradigmatic system for such investigation is represented by a two-level emitter coupled to a confined mode of the electromagnetic field \cite{JC}. The experimental benchmark of a coherent light-matter interaction is the creation of hybridized modes, which can be observed if the coupling between the field and the emitter is larger than that between these and environment. This strong-coupling regime has been achieved by employing a variety of emitters, ranging from Rydberg atoms to superconducting qubits, all characterized by large electric-dipole transition amplitudes \cite{Nori}. Spin-photon coupling is much weaker, but can be dramatically enhanced by exploiting cooperative phenomena in $N$-spin ensembles (SEs) \cite{Dicke,TC}. In this way, the strong coupling regime has been demonstrated with different spin systems in high quality factor microwave resonators \cite{Kubo2010,Amsuss,Schuster,Probst}. Along the same lines, experimental evidence of the coherent coupling between 3D-cavity photons and magnons in ferro- and ferri-magnetic crystals has been provided \cite{Huebl, Tabuchi, Zhang1}. 

Molecular spin systems display features that are potentially exploitable in quantum-information processing \cite{GTA}, such as a wide tunability of the physical parameters and decoherence times exceeding 10$^3$ the gating times at liquid nitrogen temperature \cite{Aeppli, Bader, Freedman, Sessoli}. Organic radicals provide possibly the simplest spin systems, consisting of single unpaired electrons with isotropic $g$-factors. In addition, the presence of intermolecular exchange interactions in non-diluted ensembles gives rise to exchange narrowing, which averages out the intermolecular dipolar and hyperfine interactions \cite{AbragamBleaney}. SEs of organic radicals can thus combine narrow magnetic transitions and high spin densities. For this reason, they are particularly suitable for reaching the strong coupling regime in a microwave cavity \cite{Chiorescu, Abe, APL}, while their versatility inspires the implementation of quantum gates \cite{Sato}.

Here we exploit these features in order to demonstrate the coherent coupling between distinguishable SEs. By using (3,5-dichloro-4-pyridyl)bis(2,4,6-trichlorophenyl)methyl radicals (PyBTM) \cite{Hattori14}, we first show that the strong-coupling regime is largely achieved in a broad temperature range, with values of the cooperativity reaching $4300$ at $2\,$K. This allows us to provide an experimental evidence of the coherent coupling between up to three spatially separated SEs, mediated by the cavity mode of a high $T_c$ YBCO/sapphire coplanar resonator. This capability represents a resource for the implementation of hybrid architectures and of quantum-information approaches based on the use of multiple spin-excitation modes.

Our experiment consists in measuring the transmission spectrum of a coplanar resonator in the presence of spin radical ensembles. An external magnetic field, applied in the plane of the superconducting film (Fig. \ref{1ens}), tunes the Zeeman energy of the spins relative to the energy of the cavity mode. The YBCO resonator explore wider ranges of temperature, magnetic field and power, with respect to the more conventional Nb cavities. Notably, the resonance frequency and the quality factor of the bare YBCO resonator are, at low temperature, weakly dependent on the external magnetic field \cite{APL}. The resonator is installed in a cryo-magnetic set-up \cite{supplemental} and the transmission scattering parameter is measured by means of a vector network analyzer. With the sample at the center of the resonator and at zero field, the fundamental mode has a frequency $\omega_c/2 \pi \simeq 7.7\,$GHz and a quality factor $Q \simeq 2.3 \times 10^4$. A typical incident power in our experiments is -13 dBm, corresponding to an average number of photons in the cavity of approximately $10^{12}-10^{14}$. 

The system we investigate essentially consists of $M$ ensembles of $s=1/2$ spins, coupled to a single cavity mode, which can be modeled by the Tavis-Cummings Hamiltonian \cite{TC}.
%\begin{equation}
%H_{TC} = \omega_c a^\dagger a 
%+ \sum_{k=1}^{M} [\omega_k S_{z,k} + \eta_k (a^\dagger S_{k,-} + a S_{k,+}) ],
%\end{equation}
If the number of excitations is much smaller than the number of spins in each ensemble, the spin states can be mapped onto those of a bosonic mode \cite{HP}. As a result, the Hamiltonian reads ($\hbar \equiv 1$):
\begin{equation}\label{eq06}
H = w_c a^\dagger a 
+ \sum_{k=1}^{M} [ \omega_k (b_k^\dagger b_k-S_k) + g_k (a^\dagger b_k + b_k^\dagger a)] ,
\end{equation}
where $\omega_c$ is the frequency of the cavity mode, $\omega_k$ the Zeeman splitting of a spin belonging to the $k$-th SE. Finally, $ g_k = \eta_k \sqrt{N_k} $, with $\eta_k$ the single-spin photon coupling. In order to simulate the observed spectra, we use of the input-output formalism \cite{IO,supplemental}. In particular, the transmission scattering parameter is given by:
\begin{eqnarray}\label{eq07}
S_{21}\!\!\!\! &=& \!\!\!\!\frac{\sqrt{\kappa_1\kappa_2}}{i(\omega_c\!\!-\!\!\omega)\! \!+\!\! \frac{1}{2}(\!\kappa_1\!\!+\!\kappa_2\!\!+\!\kappa_{int}\!)\!\!+\!\!\sum_{k=1}^M \!\frac{g_k^2}{i(\omega_k\!-\omega)+\gamma_k/2}},
\label{inout}
\end{eqnarray}
where $\kappa_1$ and $\kappa_2$ represent the cavity-photon escape rates to the two sides of the cavity, $\kappa_{int}$ accounts for additional cavity-relaxation channels, and $\gamma_k$ is the relaxation rate of the $k$-th spin mode.  

%%%%%%%%%%%%%%%%%%%%%%%
\begin{figure}[tbp]
\centering
\includegraphics[width=8cm]{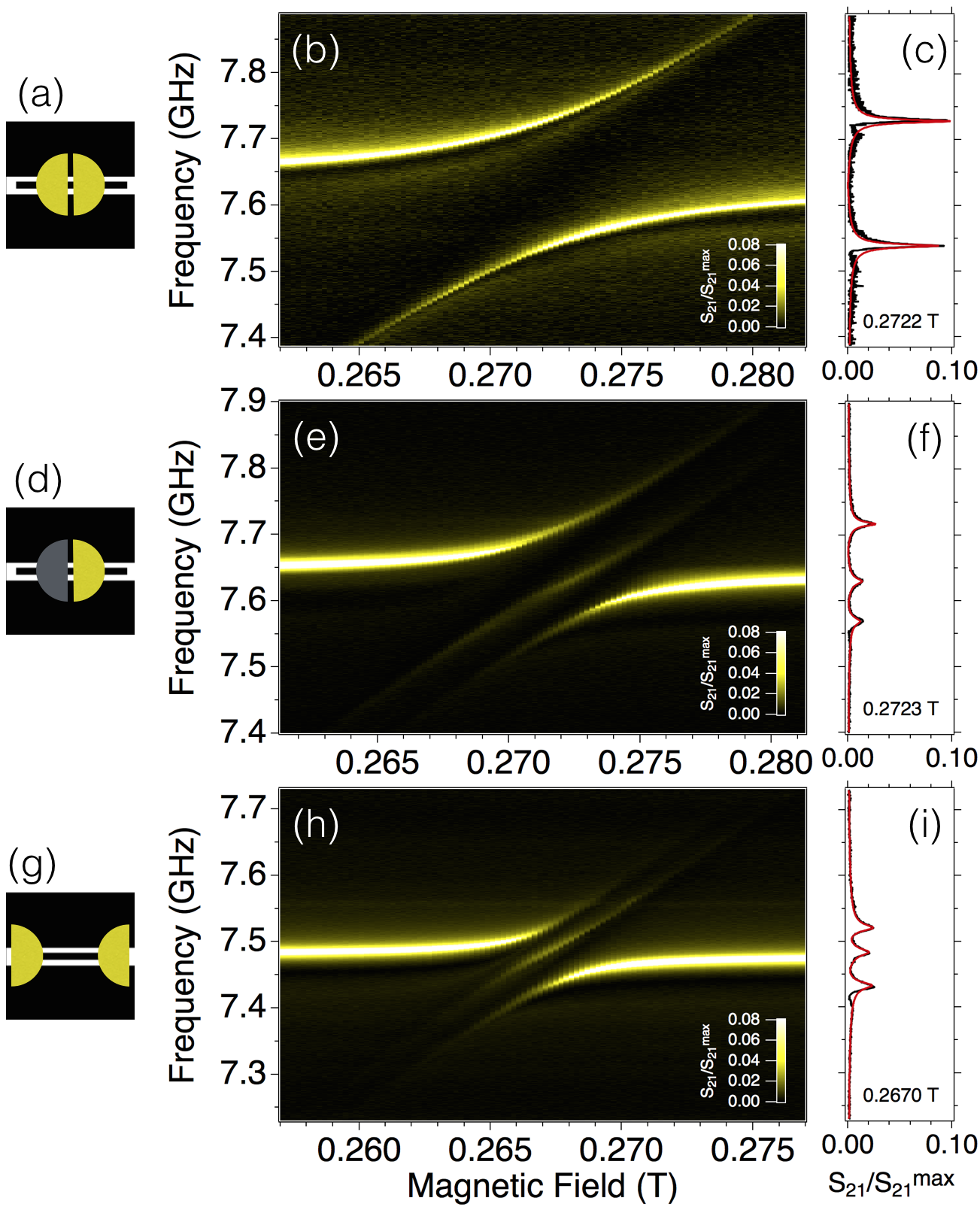}
\caption{Transmission spectra involving two SEs ($T=2\,$K): (a,b) two PyBTM samples at the center; (d,e) one PyBTM and one DPPH sample at the center; (g,h) two PyBTM samples at the boundaries of the central electrode. The panels (c,f,i) contain the comparison between experimental (black) and simulated (red) spectra at the anticrossing field.}
\label{2ens}
\end{figure}
%%%%%%%%%%%%%%%%%%%%%%%

We initially consider the case of a single PyBTM ensemble positioned at the antinode of the magnetic field, corresponding to the center of the resonator (Fig. \ref{1ens}). 
The transmission spectrum displays a well-defined anticrossing as a function of the magnetic field, with a gap of about $ 191\, $MHz [Fig. \ref{1ens}(a)]. The transmission spectra can be simulated by means of Eq. (\ref{eq07}), which is also used to estimate the Hamiltonian parameters and the relaxation rates: $g_{1}=95\,$MHz, $\gamma_1 = 7\, $MHz, and $ \kappa_1 = \kappa_2 \simeq 4\,$kHz \cite{supplemental}. The cavity-photon lifetime is limited by the internal-loss rate, $\kappa_{int} \simeq 0.3\,$MHz, estimated by fitting the transmission spectrum far off resonance \cite{supplemental}. Altogether, these parameters allow us to reproduce the Lorentzian line shapes of the peaks at resonance. Being the spin-photon coupling larger than the dissipative rates, the strong spin-photon coupling is clearly achieved, with values of the cooperativity $C = g_1^2/\gamma_1\kappa_1 $ as high as 4300 at $2\,$K. This value is two orders of magnitude larger than those typically reached with spin impurities in crystals \cite{Nori} and even larger than those obtained with exchange-locked ferrimagnetic systems \cite{Huebl}. 

%%%%%%%%%%%%%%%%%%%%%%%
\begin{figure*}[th]
\centering
\includegraphics[width=15cm]{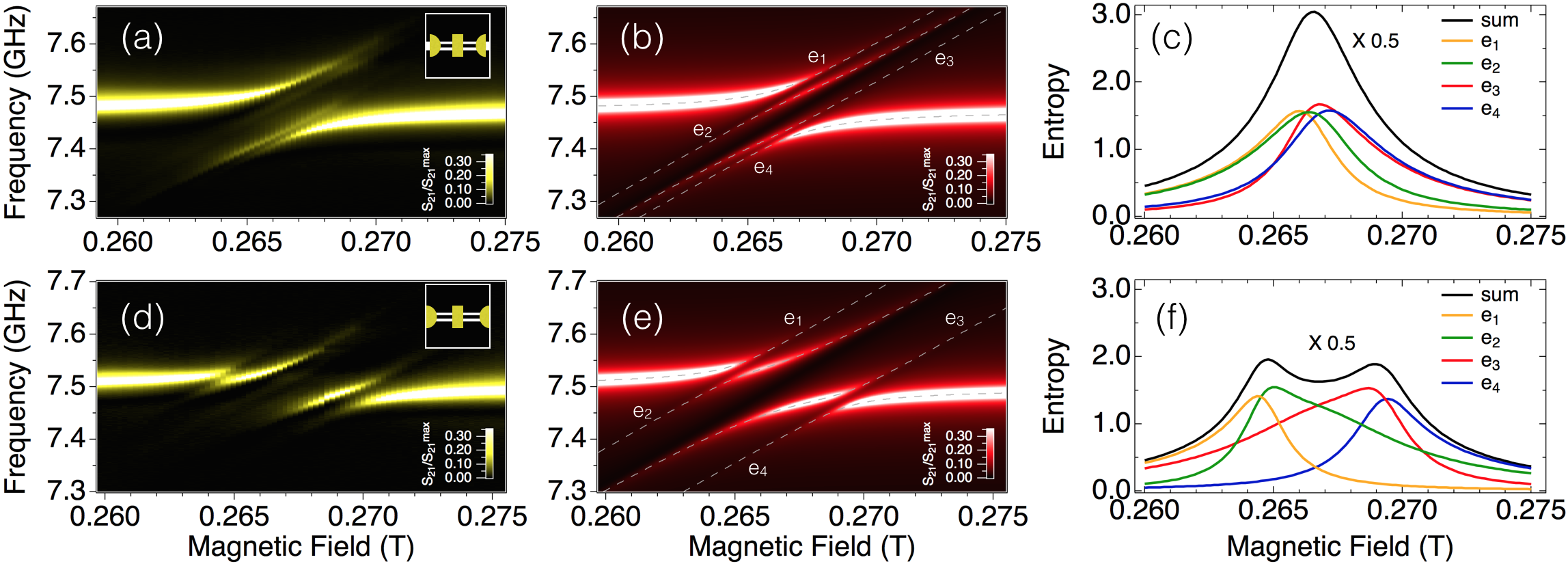}
\caption{Transmission spectra involving three SEs, one at the center of the resonator and two on the edge of the central electrode (a) or on the gap between the this and the external box (b). (b,e) Spectra simulated through the input-output formalism. The dashed lines show the Hamiltonian eigenvalues. (c,f) Entropic measures of the normal modes, $\mathcal{S}_k$ and $\mathcal{S}$, calculated as a function of the magnetic field.}
\label{3ens}
\end{figure*}
%%%%%%%%%%%%%%%%%%%%%%%

The spin-photon coupling is expected to display a temperature dependence, which can be investigated thanks to the stability of YBCO resonator in a wide temperature range \cite{APL}. The results are displayed in Fig. \ref{1ens}(b) for three different PyBTM ensembles and compared with the square root of the polarization factor 
$ p(T) = M(T) / M(0) \le 1 $
(solid lines), which are independently derived from the measurement of the magnetization $M$ \cite{supplemental}. The trends of the two quantities are indeed similar:
$ \overline{g} (T) / g(0) \simeq \sqrt{p(T)} $.
On the other hand, the temperature dependence of both the polarization factor and the magnetic susceptibility \cite{supplemental} significantly deviates from what expected for an ensemble of uncoupled spins. This suggests that a weak antiferromagnetic coupling among the spins of each ensemble is indeed present. The temperature is also expected to affect the relaxation rates of both the cavity and the spin modes. Here we find that $\gamma_1$ monotonically increases with temperature in the range $2-30\,$K, while the overall photon-decay rate remains below $ 0.5 \,$MHz. The strong-coupling regime is thus preserved at least up to $T=30\,$K. Based on the estimated value of the single-spin coupling $ \eta = 0.6\,$Hz \cite{APL}, we finally estimate the number of spins that form the three SEs, which is of the order of $10^{17}$. 

The phenomenology becomes richer when physically distinguishable SEs are simultaneously coupled to the cavity mode. Such distinguishability results either from the use of spins with different $g$-factors or from inhomogeneities in the magnetic field. In the following, we provide examples of both approaches. 
We preliminary consider the case where the above PyBTM sample is divided into two parts, positioned at the center of the resonator, at about 200$\,\mu$m from one another [Fig. \ref{2ens}(a)]. The transmission spectrum shows a single anticrossing as a function of the magnetic field, centered at the resonance field of PyBTM [Fig. \ref{2ens}(b)]. Being the two ensembles degenerate ($\omega_1 = \omega_2$), the overall system is characterized by a dark and a bright mode. The latter gives rise to a spectrum that coincides with that of a single, large PyBTM ensemble, and can be reproduced by using the previous set of parameters \cite{Kurucz, Diniz}. We then replace one of the two PyBTM ensembles with an DPPH ensemble, without modifying the system geometry, noting that the $g$-factor of DPPH ($ 2.0037 $) is slightly different from that of PyBTM ($ 2.003 $). An additional, faint line now appears between the two main ones at the anticrossing [Fig. \ref{2ens}(f)], resulting from the mixing between the bright and the dark modes discussed above. Being this effect absent in the previous case, it can be attributed to the slight difference in terms of $g$-factor between the two radicals. The transmission spectrum can be reproduced by assuming $g_{1}=50\,$MHz and $\gamma_1=8\,$MHz for PyBTM, $g_{2}=37\,$MHz and $\gamma_2=14\,$MHz for DPPH \cite{supplemental}. 

An inequivalence between two spatially separated SEs can also result from gradients in the applied magnetic field across the resonator, which here approximately corresponds to $0.9\,$T/m \cite{supplemental}. In order to exploit this effect, we position two identical ensembles of PyBTM at the opposite sides of the resonator [Fig. \ref{2ens}(g)]. The transmission spectra [Fig. \ref{2ens}(h)] qualitatively reproduce those obtained in the previous case. Here, however, being the spins that form the two ensembles identical, the physical distinguishability can only be ascribed to the magnetic-field gradient. As expected, the coupling to the resonator mode is weaker than the one observed in the previously considered geometry ($g_{1}=22\,$MHz and $g_{2}=29\,$MHz \cite{supplemental}), where all the spins are positioned at the antinode of the magnetic field. Besides, the observed line widths lead to larger estimated values of the spin relaxation rates ($\gamma_{1}=\gamma_{2}=13\,$MHz), probably due to the inhomogeneity of the static field at the electrode boundaries.

The possibility to reach the strong coupling even when the SEs are located at the boundaries of the central electrode allows us to study the coupling of three inequivalent SEs with the resonator mode. In order to observe the peculiar features of a three-SE system, the difference between the Zeeman energies of any two ensembles must be comparable to the collective spin-photon coupling \cite{supplemental}. Such condition can be met on one hand by tailoring the size of the PyBTM ensembles (and thus the spin-photon coupling) and, on the other hand, by tuning the differences between their Zeeman energies through a suitable positioning of the SEs within the superconducting electrode. In particular, if we position one SE at the center of the central electrode and two at the edges, we observe a multiple level crossing [Fig. \ref{3ens}(a,b)], with four lines in the same range of values of the magnetic field ($g_1=13\,$MHz, $g_2=20\,$MHz and $g_3=49\,$MHz). If the two lateral SEs are positioned on the gaps between the central electrode and the external box, the ratio between the difference in the Zeeman energies and the average spin-photon coupling ($g_1=12\,$MHz, $g_2=18\,$MHz and $g_3=38\,$MHz) is increased with respect to the previous case. As a result, the three SEs give rise to sequential and nearly independent anticrossings, each one involving three lines [Fig. \ref{3ens}(d,e)].

The qualitative difference between the transmission spectra obtained in the two latter geometries can be put on a quantitative basis with the aid of an entropic measure of the coherent mode mixing: 
\begin{equation}
\mathcal{S} = \sum_{k=1}^{M+1} \mathcal{S}_k = \sum_{k,l=1}^{M+1} (\mu^k_l)^2 \log_2 (\mu^k_l)^2 .
\end{equation}
Here, the entropy $\mathcal{S}_k$ thus quantifies the degree of coherent mixing of the normal mode $c_k = \mu_1^k a + \sum_{l=1}^{M} \mu_{l+1}^k b_l$ in terms of the resonator and of the spin modes, while $\mathcal{S}$ gives the overall mixing characterizing the whole set of normal modes. In the case of the first geometry, all the entropies $\mathcal{S}_k$ and $\mathcal{S}$ present a maximum at approximately the same value of the field [Fig. \ref{3ens}(c)]. In the case of the second geometry, the maxima of $\mathcal{S}_1$ and $\mathcal{S}_2$ are clearly separated from those of $\mathcal{S}_3$ and $\mathcal{S}_4$, such that $\mathcal{S}$ displays two distinct maxima [Fig. \ref{3ens}(f)]. At each of them, one of the normal modes is dark, has a vanishing small entropy and thus approximately corresponds to a bare spin mode $b_l$. We note that in both cases, the bright modes at the anticrossing achieve values of $\mathcal{S}_k$ larger than 1, which is the highest value achievable in the case of a mixing between the cavity mode and a single spin mode. This clearly shows that the observed spectral features result from a coherent coupling between spin excitations belonging to different and spatially separated ensembles.

In conclusion, we have demonstrated the coherent coupling between physically distinguishable ensembles of organic radicals. The fingerprint of such coupling is represented by the multiple anticrossing between hybridized spin-photon modes. These observations are allowed by the high cooperativity ($C \simeq 4300$) achieved in our device at low temperatures. A wider control on such coherent coupling between remote SEs can result from the generation of strong field gradients across the resonator. In this respect, the resilience to the magnetic field of the YBCO resonators offers novel opportunities. Additional possibilities for the realization of hybrid quantum architectures arise from the wide range of $g$-factors offered by molecular spin systems. 

The authors acknowledge M. D'Arienzo and M. Sassi for their help in the  preparation and  characterization of the organic sample. This work was funded by the Italian Ministry of Education and Research (MIUR) through ``Fondo Investimenti per la Ricerca di Base'' 
(FIRB) project RBFR12RPD1 and Progetto Premiale EOS, by the US AFOSR/AOARD program, 
contract FA2386-13-1-4029 and by the European FP7 FET project $MoQuaS$ contract 
N.610449.\\

\section*{APPENDIX I: EXPERIMENT}

\subsection{YBCO coplanar resonator}

%%%%%%%%%%%%%%%%%%%%%%
\begin{figure}[bpb]
\begin{center}
\includegraphics[width=8cm]{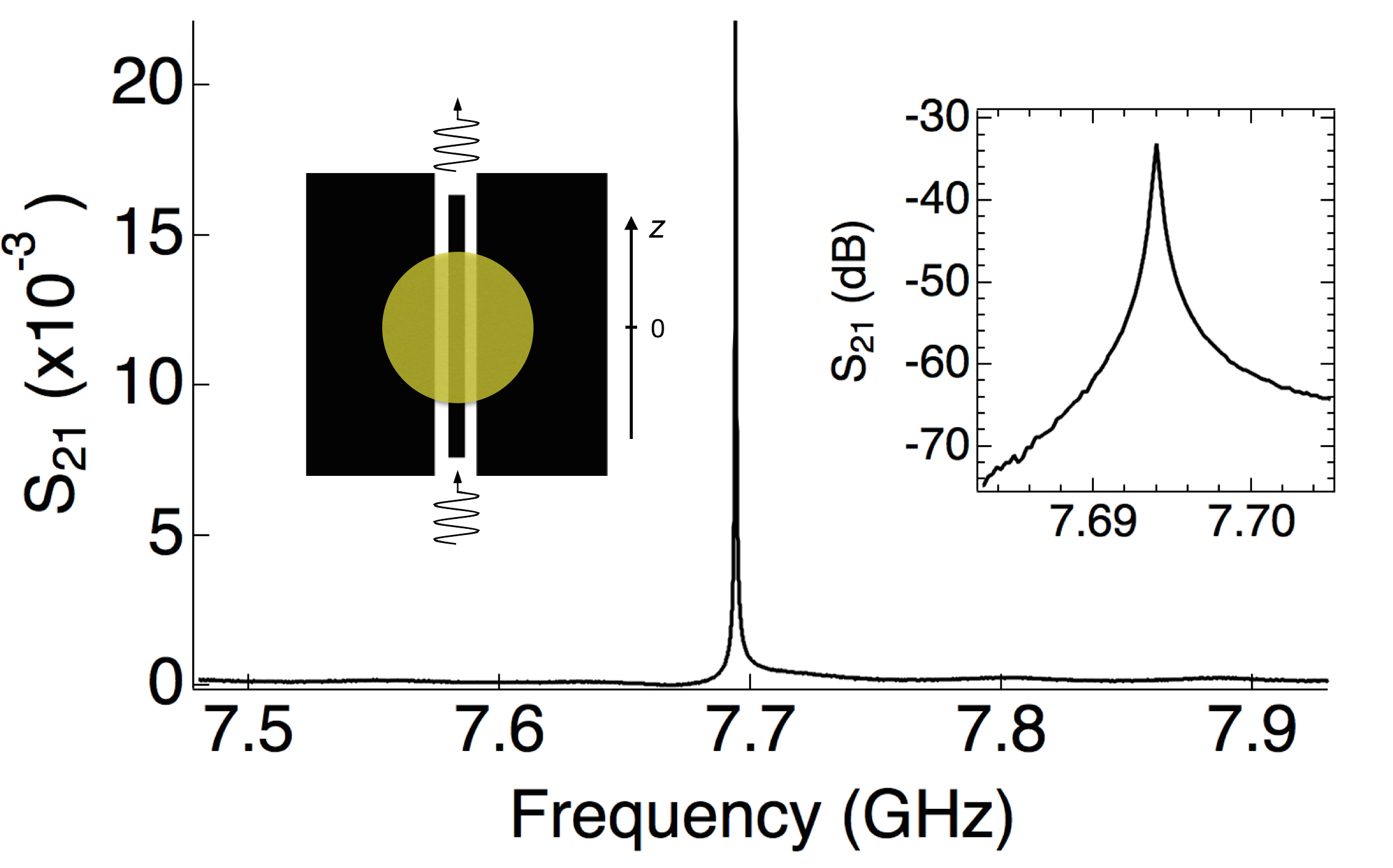}
\end{center}
\caption{Transmission spectrum taken at $2\,$K in zero magnetic field. Right inset: Transmission spectrum in dB scale. Left inset: Schematic representation of the YBCO resonator loaded with a PyBTM ensemble.}
\label{ybco}
\end{figure}
%%%%%%%%%%%%%%%%%%%%%%%

The YBa$_2$Cu$_3$O$_7$(330 nm)/sapphire coplanar resonator is fabricated by optical lithography upon wet etching of a commercial double-sided YBCO film on sapphire substrates \cite{APL}. In order to perform the measurements, the superconducting resonator is installed in a copper box and cooled down to $2\,$K in a cryo-magnetic set-up (Quantum Design PPMS $7\,$T). The transmission scattering parameter ($S_{21}$) is amplified at room temperature and acquired by means of a vector network analyzer (Agilent PNA). At $2\,$K the bare resonator shows quality factor $Q \simeq 3.37 \times 10^4$ and resonance frequency $\omega_c/2 \pi = 7.764\,$GHz.  Quality factors above $10^4$ can persist up to $ T \sim 55\,$K and in the presence of high magnetic field \cite{APL}. Fig. \ref{ybco} shows a spectrum taken in zero field with the sample loaded at the center of the resonator. The fundamental mode has $\omega_c/2 \pi \simeq 7.694\,$GHz and $Q \simeq 2.3 \times 10^4$, the latter being determined from the half-power bandwidth $\kappa/2 \pi \simeq 0.3\, $MHz. The coupling of the resonator to the feed lines can be varied in a controlled manner by finely tuning the positions of the launchers. In all the measurements presented hereafter the incident power is -13 dBm, corresponding to an average number of photons in the cavity of approximately $10^{12}-10^{14}$. 

For transmission-spectroscopy experiments, powder samples of organic radicals were pressed into thin pellets and glued on the surface of the YBCO film. Off-resonance background signals due to the coaxial line have been subtracted from the experimental data. The estimated total number of spins in each ensemble is $ N \approx 10^{17}$, which is much larger than the average number of photons that are present in the cavity during the experiment.

Close to the surface of the superconducting electrodes, the magnetic field is expected to present distortions, which are generally attributed to the Meissner effect and to flux penetration \cite{Clauss}. These inhomogeneities are discussed in the NMR and ESR literature, and typically give rise to a broadening of the magnetic-dipole transitions line and to a shift of the resonance field \cite{Rakvin}. In order to characterize the spatial variation of the applied magnetic field, we have performed a series of calibration measurements with a di(phenyl)-(2,4,6-trinitrophenyl)iminoazanium (DPPH) reference sample (Sigma-Aldrich). Initially, both the resonator and the sample were rigidly translated along the direction of the applied field ($z$ axis, see Fig. \ref{ybco}). The shift of the measured resonance field of DPPH suggests the presence of a weak gradient, which is consistent with the nominal inhomogeneity of the magnetic field in our experimental set-up ($\pm 0.001 \%$). Subsequently, we translated the DPPH sample along the $z$ direction, while keeping fixed the position of the resonator. In this case the shift of the DPPH resonance line is more substantial and suggests the presence of a field gradient of approximately $0.9\,$T/m.

\subsection{Magnetic properties of the PyBTM organic radical}

%%%%%%%%%%%%%%%%%%%%%%
\begin{figure}[bpb]
\begin{center}
\includegraphics[width=8cm]{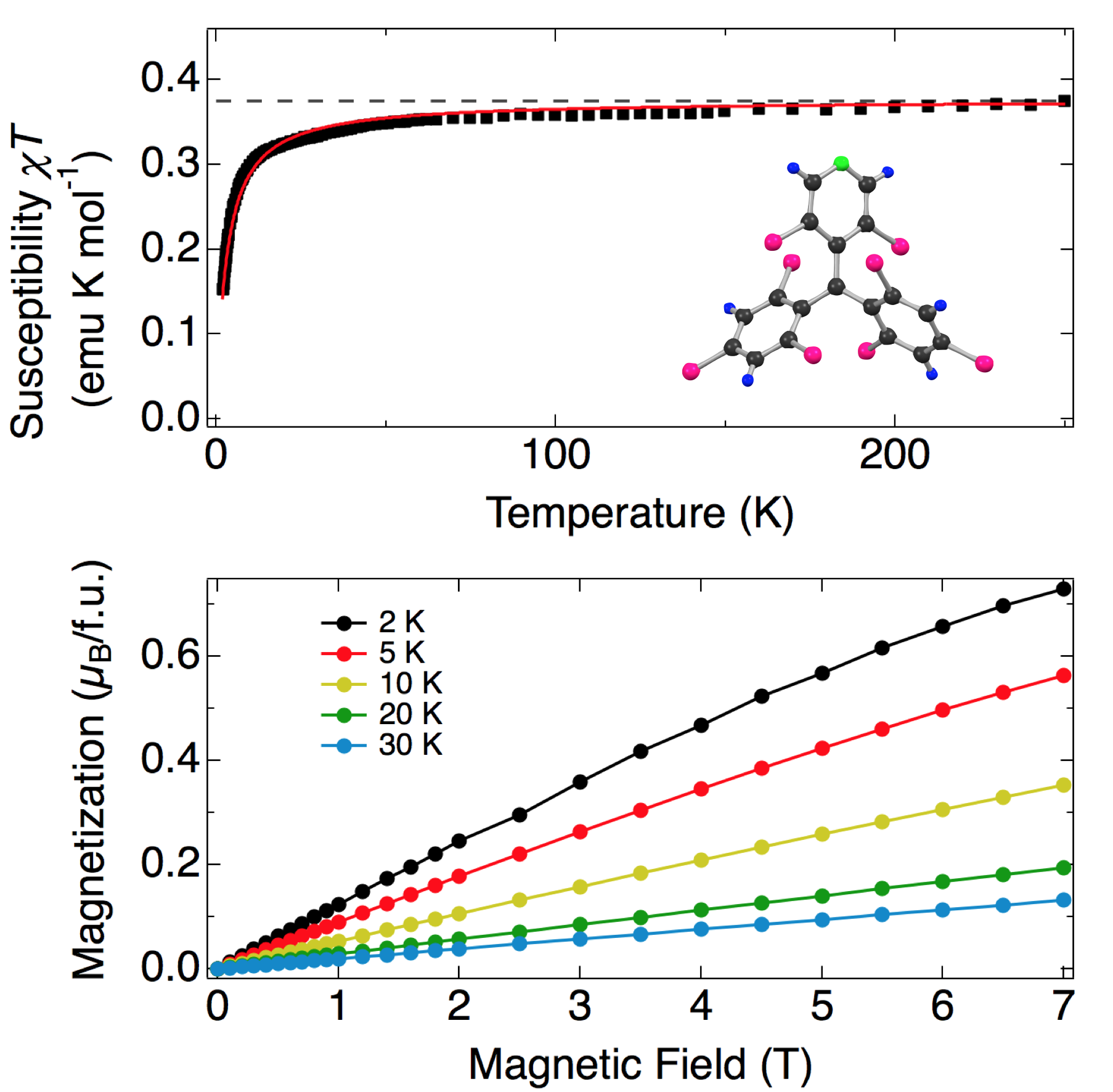}
\end{center}
\caption{Upper panel: molar susceptibility $\chi T$-vs-$T$ curve of the PyBTM organic radical. The solid line shows the fit with a Curie-Weiss model, the dashed line displays the paramagnetic limit expected for non-interacting $s=1/2$ spins. Inset: molecular structure of the PyBTM molecule (colors: C, black; Cl, magenta; N, green; H, blue). Lower panel: magnetization curves measured at different temperatures as a function of the magnetic field.}
\label{bulk_mag}
\end{figure}
%%%%%%%%%%%%%%%%%%%%%%%

PyBTM derivative was prepared according to Ref. \cite{Hattori14}. All identity and purity characterizations were in accordance with those previously reported. The powder sample of the PyBTM radical is magnetically characterized by measuring the molar susceptibility ($\chi$) and the magnetization ($M$) as a function of temperature ($T$). The behavior of $\chi T$ at high temperatures ($T\gtrsim 50\,$K) is typical of a paramagnetic system (upper panel in Fig. \ref{bulk_mag}). At lower temperatures, however, we observe a downturn of $\chi T$ with respect to the paramagnetic value (dashed line), which suggests the presence of weak intermolecular interactions with antiferromagnetic character. This hypothesis is corroborated by the quasi-linear behavior of the  magnetization curve at $2\,$K (lower panel). The temperature dependence of $\chi T$ can be reproduced by a Curie-Weiss model with $T_N=-3\,$K (solid line). 

\subsection{Additional transmission spectroscopy data}

%%%%%%%%%%%%%%%%%%%%%%
\begin{figure}[bpb]
\begin{center}
\includegraphics[width=8cm]{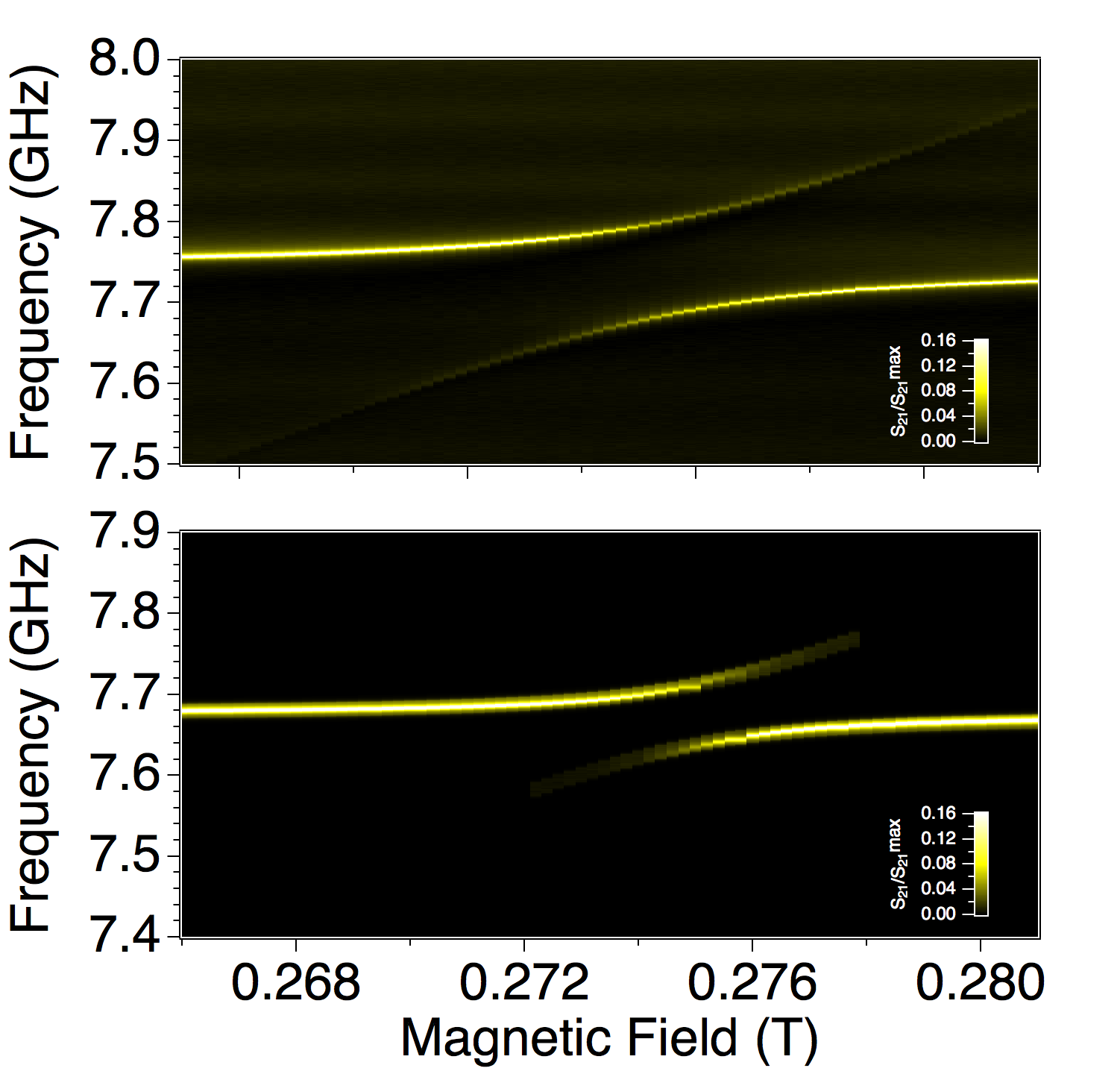}
\end{center}
\caption{Transmission spectroscopy maps measured with two ensembles of different size ($T=2\,$ K).}
\label{1ens_SI}
\end{figure}
%%%%%%%%%%%%%%%%%%%%%%%

Fig. \ref{1ens_SI} shows the transmission spectral maps measured with two ensembles of different size, which correspond to the data reported in Fig. 1 of the letter.  We estimated a total number of $2.7 \times 10^{17}$ spins for the upper panel and $0.9 \times 10^{17}$ spins for the lower panel. 

%%%%%%%%%%%%%%%%%%%%%%
\begin{figure}[bpb]
\begin{center}
\includegraphics[width=8cm]{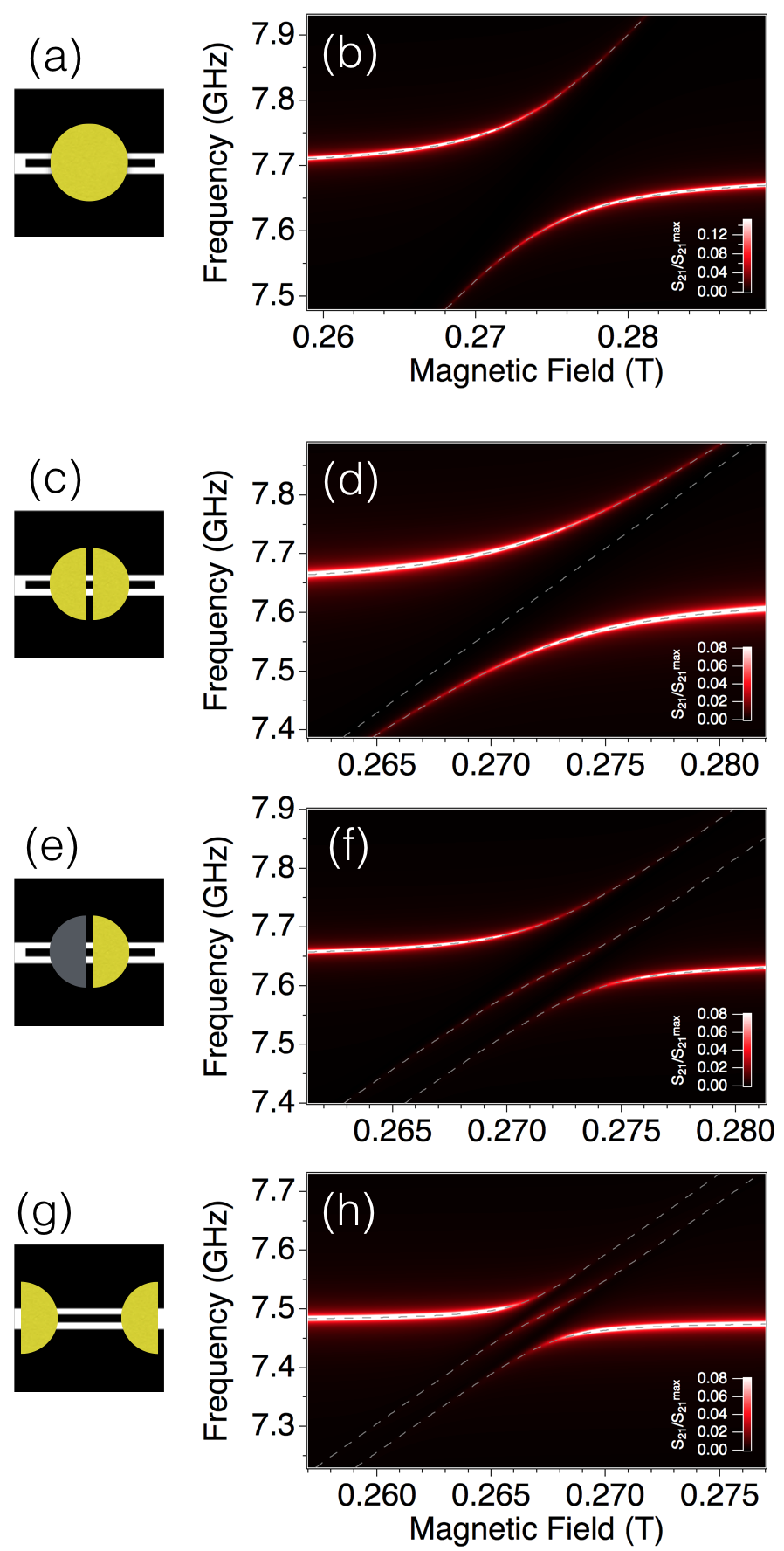}
\end{center}
\caption{Left column. Schematic of the transmission spectroscopy experiment: (a) a single PyBTM ensemble positioned at the center; (c) two PyBTM ensembles positioned at the center; (e) one PyBTM and one DDPH ensemble positioned at the center and (g) two PyBTM ensembles positioned at the edges. Right column. Input-output simulations calculated for the experiment sketched in the correspondent left panel.}
\label{IO_simulations}
\end{figure}
%%%%%%%%%%%%%%%%%%%%%%%

\section*{APPENDIX II: THEORETICAL BACKGROUND}\label{theory}

\subsection{System Hamiltonian}

We consider a system formed by $M$ ensembles of $s=1/2$ spins, coupled to a single mode of the cavity. Within the rotating-wave approximation, this is described by the Tavis-Cummings Hamiltonian ($\hbar \equiv 1$) \cite{TC}:
\begin{equation}
H_{TC} = \omega_c a^\dagger a 
+ \sum_{k=1}^{M} [\omega_k S_{z,k} + \eta_k (a^\dagger S_{k,-} + a S_{k,+}) ],
\end{equation}
where $\omega_c$ is the frequency of the cavity mode, $\omega_k$ the Zeeman splitting of a spin belonging to the $k$-th spin ensemble, and $\eta_k$ the coupling between each of such spins and the cavity mode.
In the low-excitation limit ($i.e.$ if the number of excitations is much smaller than the number of spins $N_k$ in each spin ensemble), one can introduce the Holstein-Primakoff transformation \cite{HP}, where the states of each spin ensemble are mapped onto those of a bosonic mode. As a result, the system Hamiltonian becomes:
\begin{equation}\label{eq06}
H = w_c a^\dagger a 
+ \sum_{k=1}^{M} [ \omega_k (b_k^\dagger b_k-S_k) + g_k (a^\dagger b_k + b_k^\dagger a)] ,
\end{equation}
where $ g_k = \eta_k \sqrt{2S_k} $ is the collective coupling between the spin ensemble and the cavity mode. If the temperature is much smaller than the Zeeman splittings, the total spin $S_k$ of each ensemble can be identified with its theoretical maximum ($S_k = N_k/2$).

The above bilinear Hamiltonian can be written in a diagonal form, as the sum of $M+1$ independent oscillators, each corresponding to a hybrid spin-photon mode. In order to derive the expressions of such modes, one can write the Hamiltonian in a matricial form: 
\begin{displaymath}
H = (a^\dagger, b_1^\dagger , \dots , b_{M}^\dagger ) \mathcal{A}
\left(
\begin{array}{c}
a \\ b_1 \\ \dots \\ b_{M}
\end{array}
\right) + \omega_c N ,
\end{displaymath}
where the operator $ N = a^\dagger a + \sum_{k=1}^M b_k^\dagger b_k $ counts the overall number of excitations and the $(M+1) \times (M+1)$ matrix $\mathcal{A}$ reads
\begin{displaymath}
\mathcal{A} = 
\left(
\begin{array}{cccccc}
0          & g_1      & g_2    & \dots  & g_{M-1}& g_M      \\
g_1        & \delta_1 & 0      & \dots  & 0      & 0        \\
\vdots     & \vdots   & \vdots & \vdots & \vdots & \vdots   \\
g_M        & 0        & 0      & \dots  & 0      & \delta_M
\end{array}
\right) ,
\end{displaymath}
with $\delta_k \equiv \omega_k -\omega_c$ the detuning between the $k$-th spin ensemble and the cavity mode. 
The detuning of the $k-$th spin ensemble depends on the $g$ factor of the spins and on the local value of the static magnetic field, being $ \omega_k $ the Zeeman energy.

The eigenmodes of the system can be found by diagonalizing $\mathcal{A}$, whose $M+1$ eigenvalues and eigenstates are denoted hereafter with 
$\lambda_k$ 
and 
$\vec{\mu}_k = (\mu^k_1,\dots,\mu^k_{M+1}) $, respectively. In terms of these quantities,
the Hamiltonian reads: 
\begin{equation}
H 
= \sum_{k=1}^{M+1} (\omega_c+\lambda_k) c^\dagger_k c_k
= \sum_{k=1}^{M+1} \Omega_k c^\dagger_k c_k
\end{equation}
where
$ c_k = \mu^k_1 a + \sum_{l=1}^{M} \mu^k_{l+1} b_{l}$.
An eigenstate of the above Hamiltonian thus correspond to a defined number of excitations in the each of the eigenmodes:
\begin{equation}
|n_1,\dots,n_{M+1}\rangle = \otimes_{k=1}^{M+1} \left(c_k^\dagger\right)^{n_k} | 0 \rangle .
\end{equation}

\subsection{Multiple mode mixing}

In the simplest case, where the cavity mode couples to only one spin ensemble ($M=1$), the eigenvalues of $H$ are 
\begin{equation}\label{eq01}
\lambda_k = \delta_1/2 + (-1)^k\, \sqrt{g_1^2+\delta_1^2/4}.
\end{equation}
The components of the corresponding eigenstates are given by the expressions
%\begin{eqnarray}
$
\mu^1_1 = g_1 \left(g_1^2+\lambda_1^2\right)^{-1/2} 
= - \mu^2_2
$
and
$
\mu^1_2 = \lambda_1 \left(g_1^2+\lambda_1^2\right)^{-1/2} 
= \mu^2_1 
$.
%\end{eqnarray}
In the following Section, we discuss the experimental evidence of a coherent mixing between the resonator mode and multiple spin ensembles. In order to observe such mixing, a number of conditions need to be met, concerning both the coupling to the environment and the character of the eigenmodes. With respect to the former point, the condition for resolving two modes $i$ and $j$ with neighboring values of the energy is that their linewidths are smaller than $|\Omega_i-\Omega_j|$. This somehow generalizes the condition for achieving the strong-coupling regime in the case of a single spin ensemble. In order to discuss the latter point, we introduce the normalized frequency shift
\begin{equation}
\xi_{ij} \equiv \frac{|\omega_i-\omega_j|}{\sqrt{g_i^2+g_j^2}} . 
\label{xi}
\end{equation}
For $\xi_{ij} \ll 1$, the two spin ensembles tend to form one dark and one bright mode, and thus give rise to the same spectrum as that formed by a single ensemble with $ N_i + N_j $ spins. In the opposite limit, $\xi_{ij} \gg 1$, the two spin ensembles give rise to independent anticrossings, as a function of the Zeeman energy. In the intermediate regime, where $\xi_{ij} \sim 1$, the excitations of the two spin ensembles hybridize with those of the cavity in the same range of magnetic field strength, giving rise to a multiple anticrossing. 
This characterization of the anticrossings can be put on a quantitative basis by introducing an entropic measure of the coherent mode mixing, such as 
\begin{equation}
S = \sum_{k=1}^{M+1} S_k = \sum_{k,l=1}^{M+1} (\mu^k_l)^2 \log_2 (\mu^k_l)^2 .
\end{equation}
Here, the entropy $S_k$ tells us to which extent the normal mode $c_k$ is distributed amongst the resonator mode $a$ and the spin modes $b_k$, while $S$ gives the overall degree of mixing that characterizes the normal modes. 

\subsection{Input-output relations}

In order to simulate the observed spectra, we make use of the standard input-output formalism \cite{IO}. The relations between the input and output modes of the two-sided cavity can be obtained by combining the two equations that define the boundary conditions for the cavity with the $M+1$ Heisenberg equations for the field and spin-ensemble annihilation operators.
The boundary conditions are given by
\begin{eqnarray}
a_{out} (\omega)+a_{in}(\omega) &=& \sqrt{\kappa_1}\, a(\omega)
\\
b_{out} (\omega)+b_{in}(\omega) &=& \sqrt{\kappa_2}\, a(\omega),
\end{eqnarray}
where $a_{in}$ and $a_{out}$ are the input and output modes, respectively, at one side of the cavity, while $b_{in}$ and $b_{out}$ correspond to the modes at the other side.
The Heisenberg equations for the annihilation operators read:
\begin{eqnarray}\label{eq02}
-i\omega a(\omega)\!\! & = &\!\! -i[a(\omega),H]-\frac{1}{2}(\kappa_1+\kappa_2+\kappa_{int}) a(\omega)
\nonumber\\
& + & \sqrt{\kappa_1}\, a_{in}(\omega) + \sqrt{\kappa_2}\, b_{in}(\omega)
\\ \label{eq04}
-i\omega b_k(\omega)\!\!&=&\!\!-i[b_k(\omega),H]-\frac{1}{2}\gamma_k b_k (\omega) \ (k\!=\!1,M) ,
\end{eqnarray}
where $\kappa_1$ and $\kappa_2$ represent the cavity photon escape rates to the two sides of the cavity, while $\kappa_{int}$ accounts for additional cavity-relaxation channels. The relaxation rate of the $k$-th spin mode is $\gamma_k$.  

The commutators involving the spin modes that appear in the Heisenberg equations are 
 given by:
\begin{equation}\label{eq05}
[ b_k (\omega) , H ] = \omega_k b_k(\omega) +              g_k a   (\omega) .
\end{equation}
The combination of Eqs. (\ref{eq04}) and (\ref{eq05}) allows one to establish a 
proportionality relation between the spin and the cavity modes:
\begin{equation}\label{eq03}
b_k (\omega) = \frac{-ig_k}{i(\omega_k-\omega)+\gamma_k/2}\, a(\omega) .
\end{equation}
The commutator between the Hamiltonian and the annihilation operator of the cavity mode 
reads:
\begin{equation}
[ a   (\omega) , H ] = \omega_c a  (\omega) + \sum_{k=1}^M g_k b_k (\omega) .
\end{equation}
After including this expression, and those in Eq. (\ref{eq03}), into Eq. (\ref{eq02}) one can express the annihilation operator of the cavity mode in terms of the two input modes:
\begin{equation}
a (\omega)\! = \!\frac{\sqrt{\kappa_1}\, a_{in} (\omega) + \sqrt{\kappa_2}\, b_{in} 
(\omega)}{i(\omega_c\!-\!\omega) \!+\! \frac{1}{2}(\kappa_1\!+\!\kappa_2\!+\!
\kappa_{int})+\sum_{k=1}^M \frac{g_k^2}{i(\omega_k-\omega)+\gamma_k/2}} .
\end{equation}

The above equations can be combined together in order to eliminate the system operators and express the output modes of the cavity as linear functions of the input modes alone:
\begin{displaymath}
\left[
\begin{array}{c}
a_{out} (\omega) \\
b_{out} (\omega) 
\end{array}
\right]
=
\left(
\begin{array}{cc}
S_{11} & S_{12} \\
S_{21} & S_{22} 
\end{array}
\right) 
\left[
\begin{array}{c}
a_{in} (\omega) \\
b_{in} (\omega) 
\end{array}
\right]
.
\end{displaymath}
The elements of the scattering matrix are given by
\begin{eqnarray}\label{eq07}
S_{12}\!\!\!\! &=& \!\!\!\!\frac{\sqrt{\kappa_1\kappa_2}}{i(\omega_c\!\!-\!\!\omega)\! \!+\!\! \frac{1}{2}(\!\kappa_1\!\!+\!\kappa_2\!\!+\!\kappa_{int}\!)\!\!+\!\!\sum_{k=1}^M \!\frac{g_k^2}{i(\omega_k\!-\omega)+\gamma_k/2}} 
\label{inout}
\end{eqnarray}
and 
$ S_{11}\!=\! S_{12} \sqrt{\kappa_1/\kappa_2} - 1 $. Analogous expressions can be obtained for the matrix elements $S_{21}$ and $S_{22}$, by swapping the indices of the photon escape rates $\kappa_1$ and $\kappa_2$.
%As expected, in the case where two spin ensembles $i$ and $j$ are degenerate ($\omega_i=
%\omega_j$) and relax with equal rate ($\gamma_i=\gamma_j$), their contribution to the 
%elements $S_{11}$ and $S_{12}$ coincides with that of a single spin ensemble 
%characterized by a coupling constant $\sqrt{g_i^2+g_j^2}$.

The outcome of the input-output simulations are reported in Fig. \ref{IO_simulations}. The simulated spectra correspond to the cases of one ($M=1$) and two ($M=2$) spin ensembles, that correspond to the experimental results shown in Fig. 1 and 2 of the letter and the parameters reported therein. The dashed lines display the calculated eigenvalues of the system Hamiltonian $H$.

\end{document}